# Electrowetting of a nano-suspension on a soft solid

Sumit Kumar[1], Pawan Kumar[2], Sunando DasGupta[1,3] Suman Chakraborty[1,2]

[1]Advanced Technology Development Centre, Indian Institute of Technology, Kharagpur 721302, India

[2]Department of Mechanical Engineering, Indian Institute of Technology, Kharagpur 721302, India

[3]Department of Chemical Engineering, Indian Institute of Technology, Kharagpur 721302, India

**Corresponding author:** suman@mech.iitkgp.ernet.in

**Abstract**

The wetting of solid surfaces can be manoeuvred by altering the energy balance at the interfacial region. While electric field acts favourably to spread a droplet on a rigid surface, this tendency may be significantly suppressed over soft surfaces, considering a part of the interfacial energy being utilized to deform the solid elastically by capillary forces. Here, we bring out a unique mechanism by virtue of which addition of nano-particles to the droplet brings in a favourable recovery of the electro-spreading characteristics of a soft surface, by realizing an alteration in the effective dielectric constant of the interfacial region. Our experiments further demonstrate that this mechanism ceases to be operative beyond a threshold volume fraction of the particle suspension, due to inevitable contact line pinning. We also develop a theory to explain our experimental observations. Our findings provide a non-contact mechanism for meniscus spreading and droplet control, bearing far-reaching implications in biology and engineering.

Most practical materials are intrinsically soft and deformable[1]. Wetting of such materials[2–6] has been receiving continuous attention, considering its outstanding relevance in self-organization of soft tissues[1,7], wound healing[7], metastatic spreading of cancer cells[8], functionalities of polymer gels[9], as well as practical applications in micropatterning of elastomeric surfaces[3], microfluidics[2] and inkjet printing[10]. Over the past years, the interaction of the contact line with the elastic behaviour of a deformable surface, leading to a local micros-scale formation known as wetting ridge, has been extensively investigated[3–6,11–14]. Despite a plethora of diverse applications, one fundamental property that has been continuously probed over the years is the contact angle of a liquid drop on a soft solid substrate[2,15–17].

Electrowetting on dielectric (EWOD) has emerged to be one of the most elegant and effective mechanisms of altering the state in interfacial energy of a solid surface being wetted by a droplet, manifested by a lowering of the contact angle[18–23]. In EWOD, a solid electrode is coated with a hydrophobic layer of dielectric material[10,24–27]. Simultaneously, a conducting liquid droplet is deployed as the other electrode so as to form a capacitive structure, albeit with an effective contact area with the electrode that in turn depends on the extent of spreading of the droplet on the dielectric layer[28–34]. By applying a voltage between the droplet and the insulator-covered electrode, the contact angle of the droplet can be drastically reduced as compared to the Young's angle, as attributable to a reduction in the effective solid-liquid interfacial energy[35,36]. EWOD finds extensive applications in digital microfluidics[29], biological assays[37,38], liquid papers[39], liquid lenses[40,41], to name a few.

While EWOD on ideal rigid substrates has been shown to be highly effective towards enhancing the spreading of droplets, the extent of spreading has recently been shown to be significantly arrested when experiments are conducted on a soft substrate[28,42]. However, such soft substrates are inevitable in most practical applications having biological relevance. Therefore, the quest of recovering the enhanced droplet spreading, which could have otherwise been realized had this surface been ideally rigid, turns out to be an outstanding proposition in interdisciplinary science.

Here, we report a strategy of recovering the intrinsic electro-spreading characteristics of a droplet on a soft substrate by deploying optimal suspensions of nanoparticles. We bring out the physics of this intriguing phenomenon by appealing to the energetics of the droplet-substrate interface, and arrive at a modified Young-Lippman equation capable of explaining this remarkable behaviour. Further, our results demonstrate that the recovery of spreading behaviour with the aid of nanofluid suspensions ceases to be effective beyond a threshold particle concentration, as attributable to contact line pinning.

The schematic of our experimental setup is shown in Fig. 1. The details of materials and methods are given in the Supplementary material (SM). The change in apparent contact angles of sessile droplets with and without nanosuspension atop the deformable solid with increasing applied electrical voltage is shown in Fig. 2 (a). The augmentation in the electrospreading of sessile droplet with the addition of nanoparticle is conspicuous. This observation is non-intuitive , and contrary to the spreading characteristics of pure liquid droplet under EWOD on soft dielectric that gets arrested due to the deformation near the three phase contact line[12–15, 26,28] as compared to the spreading on rigid dielectric films (Fig. 2 (b), Fig. S2 (a), SM). Our investigation further reveals that the wetting of droplet enhances under electrowetting until a threshold value of nanoparticles suspension reaches. The suppression of electrospreading of droplet beyond a threshold concentration of particles occurs due to enhanced contact line pinning.

The classical Young-Lippmann equation which is valid for conducting fluid (without colloidal suspension) on rigid surface can be written as[22]

$$\cos\theta = \cos\theta_0 + \frac{\varepsilon_0 \varepsilon_r}{2h\gamma}V^2. \qquad (1)$$

The electrowetting of droplets with nanosuspension on deformable soft dielectric may lead to the deviation from the classical Young-Lippmann theory due to nanoparticles adsorption or absorption across the droplet interfaces and electro-elasto capillary interactions. When a neutral dielectric comes in contact with an aqueous electrolyte, a charge is developed on the surface of the dielectric[43,44], which is often attributed in the negative zeta potentials ($\zeta$) shown by such systems[45] . From the electro-capillarity based approach to EWOD[46,47] , we know that the application of external electric field results in the formation of an electric double layer (EDL) at the droplet-dielectric film interface (Fig. 1 (a), Supplementary material (SM), Fig. S1 (a)). This spontaneous charge accumulation at the interface is responsible for the reduction in solid-liquid interfacial $(\gamma_{SL})$ tension[46,47]. The change in the solid-liquid interfacial tension can be written as[28]

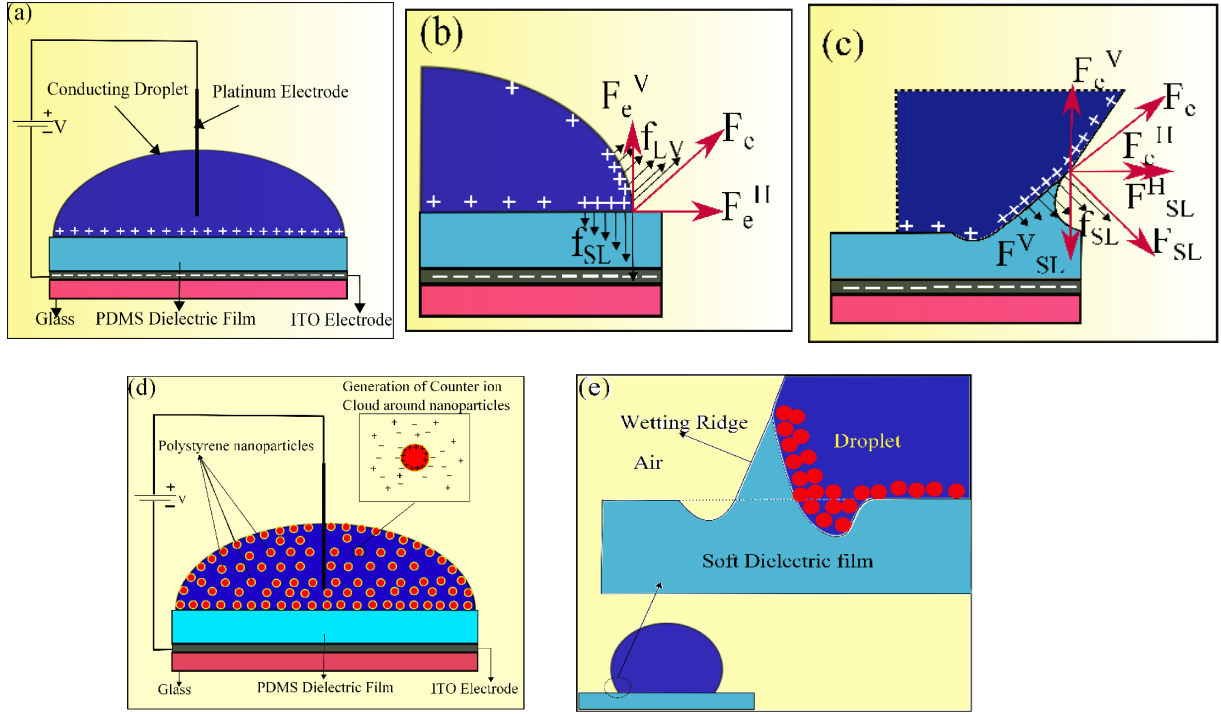

**Fig. 1 (a)** Schematic of a typical Electrowetting on Dielectrics configuration. A sessile droplet is placed on an insulating layer of some hydrophobic material (PDMS in this case). The dielectric is coated on the top of a conductor. The charges developed at droplet dielectric interface are shown. Considering the droplet to be a perfect conductor, the entire charge developed upon application of an electric potential difference shall entirely reside on the surface of the droplet (charges on the liquid-vapor interface are not shown in the figure). **(b)** Electrical forces at the Triple Phase Contact Line (TPCL) are shown. The net electrical force $\left( F_e = \dfrac{\varepsilon_0 \varepsilon_r}{2h\gamma}\left(\dfrac{V^2}{\sin\theta}\right)\right)$ on the liquid-vapor interface can be resolved into horizontal $\left( F_e^H = \dfrac{\varepsilon_0 \varepsilon_r}{2h\gamma} V^2 \right)$ and vertical $\left( F_e^V = \dfrac{\varepsilon_0 \varepsilon_r}{2h\gamma} V^2 \cot\theta \right)$ components. It is $F_e^H$ which causes the contact line movement, resulting in the spreading of droplet upon application of an external voltage $V$. The force distribution $\vec{f}_{SL}$, per unit area acting on the substrate due to Maxwell stresses at the droplet-dielectric interface is balanced by the normal reaction provided by the substrate. **(c)** Electrical forces at the Triple Phase Contact Line (TPCL) are shown for a soft substrate. The vertical component of the capillary force $(\gamma \sin\theta)$ and the electric force $(F_e^V)$ at the TPCL, creates a deformation in the substrate. The horizontal component $(F_{SL}^H)$ of the force distribution $\vec{f}_{SL}$, per unit area acting on the substrate due to Maxwell stresses at the droplet-dielectric interface contributes to a de-pinning force acting at the TPCL. **(d)** Sessile droplet containing nanoparticles in electro-wetting on a dielectric configuration. Electric double layer is generated on polystyrene nanoparticles when it comes in contact with the aqueous electrolyte of the droplet. Counter ion cloud generated around a nanoparticle has been shown in the figure. **(e)** Schematic of a droplet lying on a soft surface and the formation of a wetting ridge due to the deformation of the soft solid. Also an accumulation of nanoparticles at the wetting ridge is shown, creating a hindrance in the movement of contact line over a soft surface.

$$d\gamma_{SL} = -\rho_{SL,eff}(V)dV. \qquad (2)$$

where $\rho_{SL,eff}(V)$ is the effective surface charge density (depends upon the applied voltage (V)) at the solid-liquid interface, which is same as the net charge density of the counter ions in the EDL[47]. When polystyrene nanoparticles are added to the sessile droplet, a charge develops on the surface of the nanoparticles[17,48] (Fig. 1 (d)). The addition of nanoparticles thus causes an increase in the effective charge density of counter-ions $(\rho_{SL,eff})$. This increased counter-ion charge density causes an increase in the wetting of a droplet upon addition of nanoparticles during EWOD by causing a greater reduction in solid-liquid interfacial tension $(\gamma_{SL})$ (Eq. 2). Furthermore, it is observed that enhancing the concentration of nanoparticles in the droplet increases the slope of the curve between contact angle depression $(\cos\theta - \cos\theta_0)$ and $V^2$ for the same dielectric layer softness (E=0.02 MPa), however, at higher concentrations the slope decreases (Fig. 2 (a)). This observation could be attributed to the fact that, as the concentration of nanoparticles increases, the charge density of counter ion cloud generated by the added nanoparticles increases, thereby causing a net increase in the effective charge density of counter ions cloud at the droplet-dielectric interface[49]. Before further delving into the effect of nanoparticle concentration on the electrowetting performance, it is worthwhile to take a look at the effect of nanoparticle concentration on the wetting and pinning of the TPCL of a sessile droplet. The solid-like ordering of the nanoparticles at the TPCL[49–52] has been found to enhance the structural disjoining pressure resulting in enhanced droplet spreading[52]. The wetting of droplet increases with increase in the concentration of nanoparticles during electrowetting, however, the wetting gets arrested at higher value of nanoparticle concentration. At lower concentration, the nanoparticles provide a lubricating effect and enhances the droplet spreading[22] (Fig. 2 (a)). The reason for decrease in the spreading of droplet at higher nanoparticle concentration is mainly due to enhanced contact line pinning. We observed that the Young-Lippmann equation (Eq. 1) fails to follow the experimental observations for the nanofluid sessile droplet (Fig.2 (a)), and hence corroborates the fact that the altered solid-liquid interfacial charge density (Eq. 2) affects the electrowetting of a sessile nanofluid droplet over a soft solid.

Now, the effect of addition of nanoparticles on droplet spreading can be quantified in terms of change in chemical potential[53,54] or in the solid-liquid interfacial tension $(\gamma_{SL})$ of the sessile droplet (Supplementary Material (SM) for detail derivation). This change in solid-liquid interfacial tension $(\gamma_{SL})$ can also be seen as a manifestation of the change in counter-ion charge density in the sessile droplet by the addition of nanoparticles[49]. Representing the initial and modified counter-ion charge densities on addition of nanoparticles as ρ and ρ′ respectively, we suggest a logarithmic law for the variation in counter-ion charge density upon nanoparticle addition, similar to effective solid-liquid interfacial tension $(\gamma_{SL})$ (Supplementary material (SM), Fig. S3 and Eq. S16)

$$\rho_{SL}^{\prime,eff} = (-a\ln(\phi)+c)\rho_{SL}^{eff}. \tag{3}$$

where 'a' and 'c' are the constants. The constant 'a' is a function of surface excess of the nanoparticles. The change in effective solid-liquid interfacial tension, on applying an electric potential difference[46,47] and using Eq. (2), can now be obtained as

$$d\gamma_{SL}^{eff} = -\rho_{SL}^{\prime,eff} dV. \tag{4}$$

Using the modified value of effective solid-liquid interfacial tension from equation (4) in electrochemical approach to electrowetting[46,47] along with Young's law[1], a modified Young-Lippmann equation can be presented as

$$\cos\theta - \cos\theta_0 = (-a\ln(\phi)+c)\frac{\varepsilon_0 \varepsilon_r}{2h\gamma}V^2. \tag{5}$$

This equation can further be modified as

$$\cos\theta - \cos\theta_0 = \frac{\varepsilon_0 \varepsilon_r^*}{2h\gamma}V^2. \tag{6}$$

where,

$$\varepsilon_r^* = (-a\ln(\phi)+c)\varepsilon_r, \tag{7}$$

and $\varepsilon_r^*$ is the modified dielectric constant of the droplet-dielectric system in the presence of suspended nanoparticles. By fitting curves with the experimental data, we have obtained the values of modified dielectric constant $(\varepsilon_r^*)$ for the droplet-dielectric system. We have further validated our proposed theory with the previous existing experimental results[55] [Fig. S2 (b), Supplementary material]. This modified Young-Lippmann theory shows a good quantitative agreement with the existing experimental results[55]. The effective dielectric constant $(\varepsilon_r^*)$ is obtained by fitting the experimental data with modified equation. The effective dielectric constant $(\varepsilon_r^*)$ are 2.7 and 4.584 for water and 0.01wt% respectively.

The addition of nanoparticles to the droplet enhances the electrowetting due to enhanced effective droplet-dielectric constant, but at the same time the interaction between nanoparticles and the wetting ridge formed at the TPCL of deformable surface[3–6,11–14] cannot be neglected. The length scale of the elastocapillary induced surface deformation in dielectrics with E=0.02 is O (γ/E)~$10^{-6}$ m which is much higher than that of the surface with E=1.5 MPa (O (γ/E)~$10^{-9}$ m)[26,28]. The influence of substrate softness upon the electrowetting performance of a sessile droplet is corroborated by the fact that the curves between $h(\cos\theta-\cos\theta_0)/\varepsilon_r$ and $V^2$ for dielectrics with E=0.02 MPa (50:1 PDMS) and E=1.5 MPa (10:1 PDMS) do not merge into a single curve (Supplementary material (SM), Fig. S4) as predicted by the Young-Lippmann equation (Eq. 1).

Understandably, the Young-Lippmann equation in its classical form is unsuitable for electrowetting over a deformable surface as it does not take into consideration the effect of energy dissipated in the wetting ridge displacement as well as the effect of Maxwell stresses acting at the droplet-dielectric interface at the TPCL (Fig. 1 (c)), both of which influence electrowetting over soft substrates. Referring back to figure 2 (a), the classical Young-Lippmann equation under-predicts the change in contact angle upon application of an electric potential ($V$) for a sessile droplet containing nanoparticles. By fitting curves with the experimental data, we have obtained the values for the modified dielectric constant $(\varepsilon_r^*)$ and the pinning force $(F_p)$ for electrowetting on soft dielectrics with E=0.02 MPa and sessile droplets with nano-suspension of 0.001, 0.01 and 0.1 % w/w. The modified Young-Lippmann equation can be written as

$$\cos\theta - \cos\theta_0 = \frac{\varepsilon_0 \varepsilon_r^*}{2h\gamma}V^2 - (mV+n); V > 0. \tag{8}$$

where $m$ and $n$ are the constants given as 0.0005, -0.0161; 0.0006 -0.0234; 0.0007, -0.0263 for 0.001, 0.01 and 0.1 5 w/w respectively. The value of $\varepsilon_r^*$ was observed to be 3.29, 3.58 and 3.65 for 0.001, 0.01 and 0.1 % w/w respectively. The variation of $\varepsilon_r^*$ with the nanoparticle concentration ($\phi$) is best approximated by a logarithmic function (Supplementary material (SM), Fig. S6) as predicted by equation (7). As depicted in Fig. 2 (a), the modified Young-Lippmann equation (Eq. 8) offers a better approximation as compared to the classical Young-Lippmann equation.

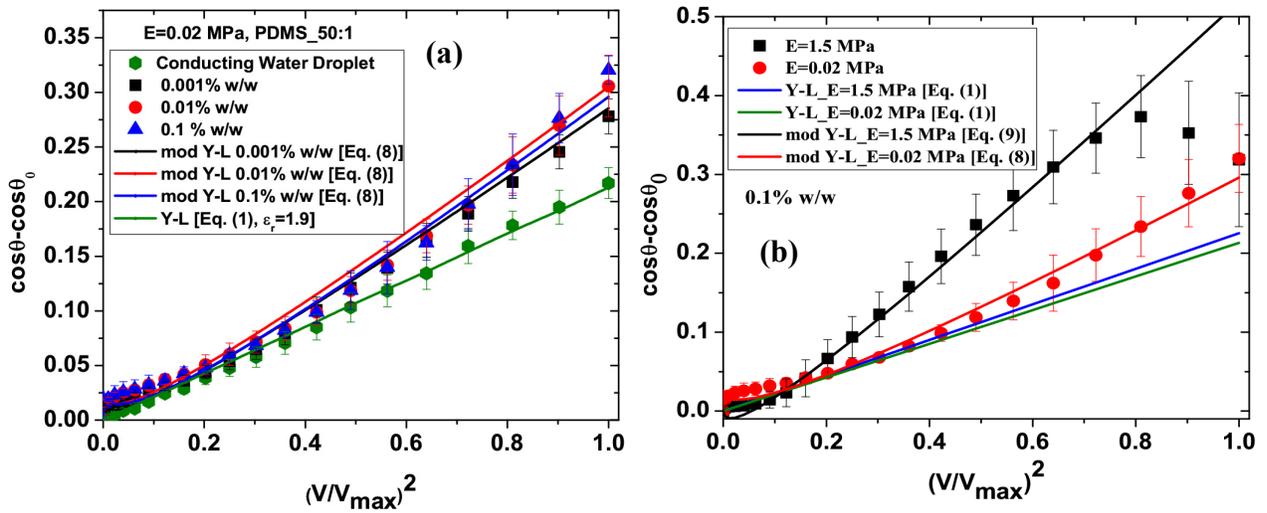

**Fig. 2** Variation of $\cos\theta - \cos\theta_0$ with non-dimensional voltage $(\overline{V} = V/V_{max})$ **(a) for deformable substrate E=0.02 MPa, with varying nanoparticles concentrations and validation of experimental results with proposed modified Young-Lippmann theory; (b) for fixed concentration of nanoparticles, 0.1 % w/w with varying elasticity modulus and validation with proposed theory**

After discussion on the electrospreading of droplets on soft solids, we here put some highlights of the droplet behavior on a non-deformable solid with varying colloidal suspension. Figure S2 (a) (Supplementary Materials) shows the modified Young-Lippmann equation (Eq. 6) and is plotted against the experimental results for 0.001 and 0.01 % w/w nanofluid droplets. The modified dielectric constant $(\varepsilon_r^*)$ was observed to be '**5.5**' and '**6.9**' for the nanoparticle concentration of 0.001 and 0.01 % w/w respectively. The modified Young-Lippmann equation (Eq. 6) is able to predict the contact angle variation with the applied voltage in a more precise manner. The modification in the effective dielectric constant of the droplet-dielectric system is corroborated by the experimental data obtained (See Supplementary Materials, Fig. S5). We reveal an enhancement in the contact line pinning during electrowetting with higher nanoparticle concentrations. Fig. S5 (Supplementary Materials, inset) shows a significant actuation voltage during electrowetting over a rigid dielectric substrate (10:1 PDMS, E=1.5) in case of 0.1% w/w nanoparticle concentration ($\approx 50\,volts$), which is almost absent in case of 0.001 and 0.01 % w/w suspension droplets. Therefore, the effect of pinning force cannot be neglected for the droplets with a higher concentration of nanoparticles like the case of soft solids. The increased pinning force at higher nanoparticle concentrations is due to increased particle-particle interactions. As the nanoparticle concentration is increased, more numbers of particles are packed in the same volume, and this upsurges the interaction amongst the particles that results in an increased pinning force. This pinning force was observed to be proportional to the applied voltage ($V$) and thus an electrostatic origin of this force cannot be ruled out.

Fig. 1 (d) shows the charges developed in a nanoparticle inside an electrolyte droplet[49]. Treating the droplet as perfectly conducting, all the charges are distributed on the surface of the droplet in the presence of an electric field. However, the few charges locked in the immobile Stern layer around the nanoparticles are not able to escape, and contribute to a net electric charge on the dielectric nanoparticle surface. This charge developed on the nanoparticles, as a result of formation of EDL on the nanoparticle surface and the redistribution of charges to the surface of conducting sessile droplet in the presence of an external electric field, gives rise to the pinning force which increases linearly with the applied voltage. In this investigation, we have introduced a linear model of pinning force depending upon the voltage ($V$) i.e. $F_p \propto V$ or $F_p = KV$, where $K$ is a constant depending upon the nanoparticle concentration in the droplet and the properties of the dielectric layer as well. With this pinning force, equation (6) can be re-written as

$$\cos\theta - \cos\theta_0 = \frac{\varepsilon_0 \varepsilon_r^*}{2h\gamma}(1 - H(V_{act} - V))V^2 - KH(V - V_{act})V. \qquad (9)$$

where, $V_{act}$ is the actuation voltage and $H$ is the Heaviside step function.

Based upon the values of $\varepsilon_r^*$ for 0.001 and 0.01 % w/w nanoparticle concentration (Equation (7)), we calculated the value of $\varepsilon_r^*$ for the nanoparticle concentration of 0.1 % w/w, which came out to be **8.29.** From the experimental data, we observed the value of $K$ to be $0.0008\gamma$ N/m (Fig. S5, Supplementary material (SM)). As evident in Fig. S2 (SM), equation (9) gives a much better prediction for the contact angle variation as compared to the classical Young-Lippmann equation (Eq. 1). The slight deviation at the higher voltages may be due to experimental errors on account of the droplet displacement observed at higher voltages. The value of coefficient $K$ shall vary with the value of actuation voltage $V_{act}$, for example, actuation voltage of 50 V ($\bar{V}=0.25$) was observed for 0.1 % w/w nanoparticle concentration ( Fig. S5, Supplementary material (SM)). This means that for the applied potential of 50 V ($\bar{V}=0.25$), the pinning forces $(F_p)$ and the electrical force in the horizontal direction $(F_e^H)$ must balance out each other, i.e. $F_p = F_e^H$, and this condition is satisfied only when $K = 0.0008\gamma$ for 0.1 % w/w nanoparticle concentration (Fig. S5, Supplementary material (SM)). The value of the coefficient $K$ can be obtained once we know the actuation voltage $(V_{act})$ for the EWOD system.

In summary, we have shown that the electro-spreading characteristics of a sessile droplet on a soft surface can be enhanced by deploying nanoparticles in the conducting fluid. This interaction is energetically tuned by altering the surface free energy with addition of suspended nanoparticles. The presence of suspended nanoparticles in the droplet, in effect, is shown to modify the counter-ion charge density of electric double layer at the droplet-liquid interface. This, on interplaying with the elastic nature of the substrate, alters the effective dielectric constant of the droplet-dielectric system. Our measured contact angle values are in quantitative agreement with the proposed modified Young-Lippmann theory.

The key insight here is that there is competing effect between the alterations in the substrate free energy due to the modified dielectric nature of the surface, which is counteracted by the enhanced contact line pinning due to interfacial particle deposition. This provides a generic mechanism that may be operative for a wide variety of electro-spreading processes relevant to objects spreading on soft media, bearing critical implications in a wide range of bio-physical and bio-chemical processes for in-vitro applications. The ability of modulating the droplet spreading using the nanoparticle and varying the effective wettability of the soft surface under imposed electrical field may find relevance in many state of art applications such as flexible optical lenses, bio-chemical detection, advanced digital microfluidics, and lab-on-a-chip devices.